\documentclass[fleqn,a4paper,12pt]{article}
\usepackage{amssymb}
\usepackage{amsmath}
\usepackage[dvips]{graphicx}
\usepackage{placeins}
\usepackage{lscape}
\usepackage{a4}
\usepackage{epsfig}

\newcommand{\ri}{{\mathrm i}}
\newcommand{\p}{\partial}
\newcommand{\bea}{\begin{array}}
\newcommand{\eea}{\end{array}}

\catcode`@=11 \long
\def\@caption#1[#2]#3{\par\addcontentsline{\csname
ext@#1\endcsname}{#1} {\protect\numberline{\csname
the#1\endcsname}{\ignorespaces #2}} \begingroup \small
\@parboxrestore \@makecaption{\csname fnum@#1\endcsname}
{\ignorespaces #3}\par \endgroup} \catcode`@=12

\newcommand{\la}{\label}

\catcode`@=11 \long
\def\@caption#1[#2]#3{\par\addcontentsline{\csname
ext@#1\endcsname}{#1} {\protect\numberline{\csname
the#1\endcsname}{\ignorespaces #2}} \begingroup \small
\@parboxrestore \@makecaption{\csname fnum@#1\endcsname}
{\ignorespaces #3}\par \endgroup} \catcode`@=12

\begin{document}

\allowdisplaybreaks
 \begin{titlepage} \vskip 2cm

%
%

\begin{center}{\Large\bf  Non-standard Dirac equations for non-standard spinors  }

\vskip 3cm {\bf {A. G. Nikitin }\footnote{E-mail:
{\tt nikitin@imath.kiev.ua} }
\vskip 5pt {\sl Institute of Mathematics, National Academy of
Sciences of Ukraine,\\ 3 Tereshchenkivs'ka Street, 01601 Kyiv, Ukraine,
\\}}\end{center}
\vskip .5cm \rm


\begin{abstract}
Generalized Dirac equation with operator mass term
is presented. Its solutions are non standard (ELKO) spinors which are eigenvectors of the charge conjugation and dual helicity
operators.
  It is demonstrated
that in spite of their non covariant nature ELKO can serve as a
carrier space of a representation of Poincar\'e group. However, the
corresponding boost generators are not manifestly covariant and
generate non-local momentum dependent transformations which are presented explicitly. These results
present a new look on group-theoretical grounds of ELKO theories.
\end{abstract}
\end{titlepage}


\section{Introduction\label{intro}}

Dark matter is seemed to be the biggest challenge for
human intellect. Being quantitative dominant substance of the
universe, it is still waiting for a consistent theoretical framework
for its description.

Till now we can indicate only  preliminary attempts to create
hypothetical elements of future dark matter theory. To create such
theory, in essence new classes of  fields are requested. One of
candidates to form the dark matter is the axion field, see, e.g.
\cite{bib}  and references cited therein. This circumstance was an
inspiration for us to analyze group-theoretical  grounds of axion
electrodynamics and construct exact solutions  for the related field
equations \cite{N1}, \cite{N11}.

Few years ago a new class of spinor fields was introduced in
\cite{Ahlu2}, \cite{Ahlu1}. They are the dual-helicity
eigenspinors of the charge conjugation operator (in German:
Eigenspinoren des Ladungskonjugationsoperators,   ELKO). The concept
of ELKO opens new interesting possibilities in constructing of
relativistic models, including the models of dark matter (see, e.g.,
refs \cite{Ahlu2}-\cite{B3}), and of other cosmological
phenomena. In particular, this concept was used to provide a new
explanation of the accelerated expansion of the universe
\cite{Ba}, \cite{B4}, \cite{Fab}. Higher dimension
aspects of ELKO theory were considered in refs \cite{liu} and
\cite{nieto} . We will not discuss the validity and
perspectives of all these models, but restrict ourselves to their
kinematical grounds connected with using ELKO.

Mathematically,  ELKO  were put to one of non-equivalent classes of
bispinor fields classified by Lounesto \cite{loun}.
 Namely, they were classified as so-called flagpole spinor fields \cite{roch}. There
 exist a clear representation of these spinors proposed in  \cite{Ahlu2}
 and \cite{Ahlu1}. Moreover, as it was indicated in
 \cite{Ahlu2} and \cite{Ahlu1},  such
 spinors satisfy the  sixteen component Dirac equation
 supplemented by two additional conditions.

 By construction, ELKO  are eigenvectors of the dual helicity
operator. Such property is not evidently compatible with Lorentz
invariance, since this operator is not a relativistic scalar for
massive fields. Moreover, in fact  ELKO contain a hidden preferred
direction that breaks Lorentz symmetry \cite{crash}.

 A natural
question arises  whether it is possible to formulate the kinematical
grounds of ELKO  theory in a more compact and relativistic invariant
manner. In paper \cite{Shaw} a manifestly covariant
generalization of ELKO concept is proposed. The related "dark matter
spinors" solve a second order field equations supplemented by
nonlocal constraints.

 Let us stress that ELKO are only subordinate constructive elements whose
 connection with  physical particle states is realized via quantum field
 operators. Moreover, the corresponding field equation should be the
 Klein-Gordon equation, but not the Dirac one, see \cite{Ahlu2},
 \cite{Ahlu1} and  paper \cite{Lie} where the most recent
 progress on the subject is presented. In other words, the sixteen component
 Dirac equation for ELKO  indicated in \cite{Ahlu2},
 \cite{Ahlu1} is not applied  to describe the dynamics of the
 corresponding fields.

On the other hand, since ELKO are used as expansion coefficients
of quantum fields, all properties of these spinors are very
interesting and important since they form
 the grounds of the corresponding field theories. By construction,
 ELKO satisfies both the Klein-Gordon  and (generalized) Dirac
 equations, and this property should be kept under a transition to
 a new inertial frame of reference. Thus there are well grounded reasons
 to study  exactly these aspects of ELKO theories, and it was the
 main motivation for writing the present paper.

 We will analyze only kinematical aspects of ELKO without refereing to the
 corresponding dynamical (field) theories. In other words our research is
 reduced to studying ELKO spinors. 

In the present paper a simple way to describe ELKO   by a
four-component generalized Dirac equation is presented. More
exactly, a  modernized Dirac equation will be used, the mass term of
which is not proportional to the unit matrix. In addition, a direct
and straightforward connection between ELKO and Dirac spinors will be
demonstrated.

We also find the explicit form of  generators of Poincar\'e group
which can be realized on ELKO. It appears that these generators, like
generators of Wigner rotations \cite{Wig}, do not have a
manifestly covariant form and generate momentum dependent and so
non-local transformations of vectors from their carrier space. These
transformations are given explicitly in the present paper. And that
is a message that ELKO theory can be treated as Poincar\'e invariant
in spite of that it is not manifestly covariant.

Finally, we present a toy model which, being transparently
relativistic invariant,  is characterized by the same kinematical
equations as ELKO.

\section{Multi-component  Dirac equation for ELKO}

Let us start with the main definitions of  ELKO theory. To save a
room we will use compact notations presented in what follows.

Like the Dirac spinors, the ELKO $\lambda^S_{\{+,-\}}=\psi^+_-, \ \
\lambda^S_{\{-,+\}}=\psi^+_+,\ \lambda^A_{\{+,-\}}=\psi^-_+ $ and
$\lambda^A_{\{-,+\}}=\psi^-_-$ satisfy the Klein-Gordon equation.
However, they do not satisfy Dirac equation, which is changed to the
following system in the momentum representation \cite{Ahlu2},
\cite{Ahlu1}:
\begin{gather}\la{1}\begin{split}&\gamma^\mu p_\mu\psi^+_++im\psi^+_-=0,\\&
\gamma^\mu p_\mu\psi^+_--im\psi^+_+=0,\\&\gamma^\mu p_\mu\psi^-_+ -im\psi^-_-=0,\\
&\gamma^\mu p_\mu\psi^-_-+im\psi^-_+=0.\end{split}\end{gather}

Like in \cite{Ahlu2} and  \cite{Ahlu1}  we use the Weyl
representation of Dirac matrices with diagonal and hermitian matrix
$\gamma_5=i\gamma_0\gamma_1\gamma_2\gamma_3$. Then ELKO  can be
specified in the following way:
\begin{gather}\la{elko}\psi^\varepsilon_\nu({\bf p})=\sqrt{\frac{E+m}m}
\left(1-\nu\frac{p}{E+m}\right)\lambda^\varepsilon_\nu\end{gather}
where
\begin{gather*}\lambda^\varepsilon_\nu=\begin{pmatrix}\varepsilon \sigma_2
\phi_\nu(0)^*\\\phi_\nu(0)\end{pmatrix},\end{gather*} $\varepsilon,
\nu=\pm,\ E=\sqrt{p^2+m^2}, \ p^2=p_1^2+p_2^2+p_3^2, \ \sigma_2$ is
the Pauli matrix, and \cite{Ahlu1}
\begin{gather*}\phi_+(0)=\sqrt{m}\begin{pmatrix}\cos\left(\frac\theta2\right)
\mathrm{e}^{-i\frac\varphi2}\\ \sin\left(\frac\theta2\right)
\mathrm{e}^{i\frac\varphi2}\end{pmatrix}, \\
\phi_-(0)=\sqrt{m}\begin{pmatrix}\sin\left(\frac\theta2\right)
\mathrm{e}^{-i\frac\varphi2}\\-\cos\left(\frac\theta2\right)
\mathrm{e}^{i\frac\varphi2}\end{pmatrix}\end{gather*}
 where $\theta$ and $\varphi$ are the polar and azimuthal angles of vector $\bf p$.

By construction, the four-component bispinors $
\psi^\varepsilon_\nu$   satisfy the following conditions:
\begin{gather}\la{2}C\psi^\varepsilon_\nu=
\varepsilon\psi^\varepsilon_\nu\end{gather} and
\begin{gather}\la{2a}\sigma_p\psi^\varepsilon_\nu=
\nu\psi^\varepsilon_\nu
 \end{gather}
where $C=\gamma_2\kappa$ is the charge conjugation operator,
$\kappa\psi=\psi^*$, and $\sigma_p=\frac1p \gamma_0\gamma_a
p_a\equiv\gamma_5
\frac{{\mbox{\boldmath$\sigma$\unboldmath}}\cdot{\bf p}}p$ is a
product of the helicity operator with matrix $\gamma_5$. In other
words, $ \psi^\varepsilon_\nu $ are eigenvectors of commuting
operators $C$ and $\sigma_p$, and just equations (\ref{1}),
(\ref{2}) and (\ref{2a}) can be used as a formal definition of ELKO.
Notice that such definition is universal and does not depend on
concrete realizations of $\gamma$-matrices and spinor components.

Relations (\ref{1})  specify a sixteen  component Dirac equation in
momentum representation for
$\Psi=\text{column}(\psi^+_+,\psi^+_-,\psi^-_+,\psi^-_-)$, and can
be rewritten as:
\begin{gather}(\Gamma^\mu p_\mu-m)\Psi=0\la{3}\end{gather}
where
\[\Gamma^\mu=\begin{pmatrix}0&-i\gamma^\mu&0&0\\i\gamma^\mu&0&0&0\\
0&0&0&i\gamma^\mu\\0&0&-i\gamma^\mu&0\end{pmatrix}\] are the
$16\times16$ Dirac matrices. Moreover, this equation should be
considered together with the additional constraints (\ref{2}) and
(\ref{2a}) which reduce the number of independent components of
$\Psi$ to 4.

Equations (\ref{3}) have a symmetric form which is  transparently
relativistic invariant. The same is true for equation (\ref{2}).
However, it is not the case for the constraint (\ref{5}) which is
not relativistic invariant in the generally accepted meaning.
Namely, covariant equations (\ref{1}) and (\ref{3}) connect modes
$\psi^\varepsilon_+ $ and $\psi^\varepsilon_- $ that are not defined
in a covariant manner.

A natural question arises whether ELKO can be treated as a
relativistic substance et all. We will see that it is the case since
they form a carrier space for a representation of Poincar\'e group.
However, this representation is not manifestly covariant.

  One more inspiration to examine the
 ELKO definition is the general feeling that there are too many
equations for a spinor with four independent components. It is
naturally  to look for a more compact kinematic equation. Just such
equation, which also opens a way to give a possible interpretation
of Poincar\'e invariance of ELKO theory, is presented in the
following section.

\section{Four-component  equation for ELKO }

Let us consider the following generalized Dirac equation
\begin{gather}\la{4}\left(\gamma^\mu p_\mu+Im\right)\psi=0\end{gather}
where $I$ is an involution which commutes  with $\gamma^\mu p_\mu$,
or
 pseudo involution anticommuting with $\gamma^\mu p_\mu$.
 Solutions of any equation of type (\ref{4}) satisfy the condition
\begin{gather}\la{7}(p_0^2-{\bf p}^2-m^2)\psi=0\end{gather}
which generate the relativistic dispersion relation. If, in
addition, $I$ commutes with generators of Poincar\'e group, equation
(\ref{4}) is transparently relativistic invariant.

 For $I$ being the unity operator equation (\ref{4}) is reduced to the standard Dirac equation. Choosing $I=\gamma_5$ we obtain a good relativistic equation which is equivalent to the Dirac one.

But there are more  involutions $I$ which satisfy the enumerated
criteria. In particular, they can be constructed using matrix
$\gamma_5$, space inversion $P$, time reflection $T$,  charge
conjugation $C$ and their products.

Let us consider a more sophisticated example of equation (\ref{4})
with $I=iC\sigma_p$:
\begin{gather}\la{5}\left(\gamma^\mu p_\mu+iC\sigma_p m\right)\psi=0.
\end{gather}
{\it Just this equation can be used to describe the kinematics of
ELKO }. Indeed, this compact expression is completely equivalent to
the cumbersome system (\ref{1}), (\ref{2}), (\ref{2a}). To prove
this statement let us introduce the following operator
\begin{gather}\la{P}P^\varepsilon_\nu= \frac14(1+\varepsilon
C)(1+\nu\sigma_p)\end{gather} where $\varepsilon$ and $\nu$ are
parameters which independently take the values $\pm1$. Since $C$
commutes with $\sigma_p$, and $C^2=\sigma_p^2=1$ operators (\ref{P})
satisfy the relations $P^\varepsilon_\nu
P^{\varepsilon'}_{\nu'}=\delta^{\varepsilon\varepsilon'}
\delta_{\nu\nu'}P^\varepsilon_\nu$ and so are projectors.

Acting on (\ref{5}) from the left by $P^\varepsilon_\nu$
 and using
the following identities:
\[\begin{split}&  P^\varepsilon_\nu\gamma^\mu p_\mu=\gamma^\mu p_\mu
P^{-\varepsilon}_{-\nu},\quad P^\varepsilon_\nu iC\sigma_p=
i\varepsilon \lambda P^{-\varepsilon}_\nu,\end{split}\] and
notations
\begin{gather}\la{6}P^\varepsilon_\nu\psi=\psi^\varepsilon_\nu. \end{gather}
we immediately come to equations (\ref{1}), (\ref{2}) and
(\ref{2a}). On the other hand, summing up all equations included
into system  (\ref{1}) and using definitions (\ref{6}) we come to
equation (\ref{5}). Thus the system (\ref{1}), (\ref{2}), (\ref{2a})
admits rather compact formulation (\ref{5}).

Let us show that equation (\ref{5}) is mathematically equivalent to
the Dirac equation. Indeed, multiplying (\ref{5}) by $\gamma_0$ we
transform  it to the Schr\"odinger form:
\begin{gather}\la{10}p_0\psi=H\psi,\quad H=\gamma_0\gamma_ap_a+
i\gamma_0C\sigma_pm.\end{gather}
Then, making the transformation
\begin{gather}\la{25}\psi\to \psi_D=U\psi, \ H\to H_D=UHU^{-1}\end{gather}
with
\begin{gather}\la{11}\begin{split}&U=\frac12(1-\gamma_5C\sigma_p)
(1-i\gamma_5),\\&
U^{-1}=\frac12(1+i\gamma_5)(1+\gamma_5C\sigma_p)\end{split}\end{gather}
we reduce (\ref{10}) to the standard Dirac equation:
\begin{gather}\la{12}p_0\psi_D=H_D\psi_D,\quad H_D=\gamma_0\gamma_ap_a+\gamma_0m.\end{gather}

Using transformation (\ref{25}) we can specify ELKO in Dirac
representation. Namely, this transformation reduces operators $C$
and $\sigma_p$ to the following form:
\[C\to UCU^{-1}=-\frac{{\mbox{\boldmath$\sigma$\unboldmath}}\cdot{\bf
p}}p,\quad \sigma_p\to
U\sigma_pU^{-1}=\gamma_5\frac{{\mbox{\boldmath$\sigma$\unboldmath}}\cdot{\bf
p}}p\] and so the counterparts of  ELKO satisfying (\ref{2}) and
(\ref{2a}) are the Dirac spinors
$(\psi_{D})_\nu^\varepsilon=U\psi^\varepsilon_{\nu}$  which are
eigenvectors of chirality operator
$\frac{{\mbox{\boldmath$\sigma$\unboldmath}}\cdot{\bf p}}p$ and
matrix $\gamma_5$, which satisfy the following relations:
\begin{gather*}\frac{{\mbox{\boldmath$\sigma$\unboldmath}}\cdot{\bf
p}}p(\psi_{D})_\nu^\varepsilon=-\varepsilon(\psi_{D})_\nu^\varepsilon,\quad
\gamma_5(\psi_{D})_\nu^\varepsilon=-\nu\varepsilon(\psi_{D})_\nu^\varepsilon.\end{gather*}

 Operator (\ref{11})
satisfies the condition $UU^\dag=1$ and includes the complex
conjugation operation. Following Wigner \cite{wig} we classify
(\ref{25}) as an unitary-antiunitary transformation.

\section{Relativistic invariance}
Equation (\ref{5}) is completely equivalent to system (\ref{1}),
(\ref{2}), (\ref{2a}) and generates a relativistic dispersion
relation (\ref{7}).  However,  the multiplier for the mass term in
(\ref{5}) is not a relativistic scalar, and this fact can be treated
as a direct proof that ELKO are not well defined covariant spinors.
On the other hand,  equation (\ref{5}) is equivalent to the
relativistic Dirac equation, and so it has to inherit its symmetries
at least in some more generalized meaning.

In the wide sense, relativistic  invariance of a differential
equation means that its solutions form a carries space of a
representation of Poincar\'e group. Let us show that equation
(\ref{5}) satisfies this weak invariance condition.

 It is a common knowledge that the Dirac
equation is relativistic invariant. Moreover, the corresponding
generators of Poincar\'e group can be represented in the following
form:
\begin{gather}\la{11a}\begin{split}&P_0= H_D,\quad P_a= p_a, \\
& J_{ab}=p_b\frac{\p}{\p p_a}-p_a\frac{\p}{\p p_b}+S_{ab},\\& J_{0a}=-p_0\frac{\p}{\p p_a}+S_{0a} \end{split}\end{gather}
where $p_0=\pm\sqrt{p^2+m^2}$ and $
S_{\mu\nu}=\frac{1}4(\gamma_\mu\gamma_\nu-\gamma_\nu\gamma_\mu).$

On the set of solutions of equation (\ref{12}) the boost generators
$J_{0a}$ can be rewritten in the following form:
\begin{gather}\la{26} J_{0a}=-\frac{1}2\left(H_D\hat x_a+
 \hat x_aH_D\right)\end{gather}
where $\hat x_a=\frac{\p}{\p p_a}-\frac{p_a}{p^2}$.

 To find a
realization of these generators on the set of solutions of equation
(\ref{5}) it is sufficient to make the transformation $P_\mu\to
P'_\mu=U^{-1}P_\mu U, \ J_{\mu\nu}\to J'_{\mu\nu}=U^{-1}J_{\mu\nu}U$
where $P_\mu$ and $J_{\mu\nu}$ with \ $\mu, \nu=0,1,2,3$ are
generators (\ref{11a}), and $U$ is operator (\ref{11}). As a result
we obtain:
\begin{gather}\la{14}P_0= H,\quad P_a=
p_a,\ J'_{ab}=J_{ab},\\\la{14a}J'_{0a}= -\frac{1}2\left(\hat x_aH
+H\hat x_a\right)+
\frac{m}{p^2}\varepsilon_{abc}\gamma_bp_c\end{gather} where $H$ is
the hamiltonian fixed in (\ref{10}).

Alternatively, starting with realization (\ref{11a}) for $J_{0a}$ and using the identities
\begin{gather}\begin{split}\la{Ka}& K_a=-U^{-1}p_0\frac{\p}{\p p_a}U=-p_0\left(\frac{\p}{\p p_a}+
\frac{\ri}{p^2}S_{ab}p_b\left(\gamma_5\sigma_pC-1\right)\right),\\&{\hat
S}_{0a}= U^{-1}S_{0a}U=S_{0a}-\frac{\ri}{p}S_{ab}p_b\left(\sigma_p
-\gamma_5C\right)\end{split}\end{gather} we obtain:
\begin{gather}\begin{split}&\la{n1}J'_{0a}=K_a+\hat S_{0a}=
-p_0\frac{\p}{\p p_a}+\Sigma_{0a}\end{split}\end{gather} where
\begin{gather}\la{n2}\Sigma_{0a}=S_{0a}+\frac{\ri}{p^2}S_{ab}p_b\left(p_0-
\gamma_0\gamma_ap_a\right)\left(1-\gamma_5\sigma_pC\right).\end{gather}

Operators (\ref{14}), (\ref{14a}) and (\ref{14}), (\ref{n1}) satisfy the following commutation relations
\begin{gather*}[P'_\mu,P'_\nu]=0,\quad
[P'_\mu,J'_{\lambda\sigma}=g_{\mu\lambda}
P'_\sigma-g_{\mu\sigma}P'_\lambda,\\
[J'_{\mu\nu},J'_{\lambda\sigma}=
g_{\mu\sigma}J'_{\nu\lambda}+g_{\nu\lambda}J'_{\mu\sigma}-
g_{\mu\lambda}J'_{\nu\sigma}-g_{\nu\sigma}J'_{\mu\lambda}\end{gather*}
which specify the Lie algebra of Poincar\'e group.

Thus we find the explicit form of generators of Poincar\'e group
which can be defined on the set of solutions of equation (\ref{10})
for ELKO. As was expected, the angular momentum operators for ELKO and
Dirac fields have  the same form.

The boost generators $J_{0a}$ and $J'_{0a}$ are different. Moreover,
they are qualitatively different. Indeed, using equation (\ref{12})
generators $J_{0a}$ can be rewritten in covariant form (\ref{11a}),
whereas generators $J'_{0a}$ do not keep this property.

\section{Lorentz transformations for ELKO}
An important quality  of realization (\ref{11a}) is that the matrix
term $S_{0a}$ which generates transformations for the wave function
commutes with the term $p_0\frac{\p}{\p p_a}$ responsible for
transformations of independent variables. As a result  Lorentz
transformations for Dirac spinors have the following generic form:
\begin{gather}\la{24a}\psi(\tilde p)\to D(\Lambda^{-1})\psi(\Lambda \tilde p)\end{gather}
where $\tilde p=(p_0, p_1, p_2, p_3),\ $ $\Lambda$ is the Lorentz
transformation matrix and $D(\Lambda^{-1})$ is a numeric matrix
dependent on transformation parameters. In particular, for Lorentz
boost we have
\begin{gather}\la{30}\begin{split}&D(\Lambda^{-1})=\exp(S_{0a}\theta_a)=\cosh\left(
\frac\theta2\right)+\frac{2S_{0a}\theta_a}\theta\sinh\left(
\frac\theta2\right)\end{split}\end{gather} where $\theta_a$ with
$a=1,2,3$ are transformation parameters and
$\theta=\sqrt{\theta_1^2+\theta_2^2+\theta_3^2}.$

 Notice that transformations for the wave function are
the same for all values of independent variables.

 The boost generator (\ref{n1})  does not have a covariant form,
and the matrix term $\Sigma_{0a}$ is much more complicated than term
$S_{0a}$ present in (\ref{11a}). It depends on $\bf p$, does not
commute with $p_0\frac{\p}{\p p_a}$ and generates dependent on $\bf
p$ transformations for $\psi$. Nevertheless, integrating the Lie
equations generated by operators (\ref{n1}), it is possible to find
Lorentz boost transformations for ELKO.

Let us note that in spite of its non-covariant form,  boost
generator (\ref{n1}) gives rise to covariant transformations
(\ref{24a}), (\ref{30})
 provided the new inertial reference frame moves parallel to momentum
 $\bf p$. Indeed, in this case the transformation parameter vector
 $\mbox{\boldmath$\theta$\unboldmath}=(\theta_1, \theta_2, \theta_3) $
 can be represented as $\mbox{\boldmath$\theta$\unboldmath}=\alpha\bf n$
 where ${\bf n}=\frac{\bf p}p$, and so $\Sigma_a\theta_a\equiv
 S_{0a}\theta_a$, exactly as in the case of Dirac equation.
 Thus it is possible to make standard transformations to the rest frame,
 study VSR aspects of ELKO \cite{Hor}, etc, etc.

But it is interesting to find generic Lorentz boost for ELKO when
transformation parameters are not dependent on momenta.
 First we note that using (\ref{n1}), it is possible to write the
infinitesimal Lorentz boost  in the following form:
\begin{gather*}\la{tr1} p_a\to p_a'=
(1-E\frac{\p}{\p p_b}\theta_b)p_a=p_a-\theta_ap_0,\\
p_0\to p_0'=(1-E\frac{\p}{\p p_b}\theta_b)E=p_0-\theta_ap_a,
\\\psi\to \psi'=(1+\Sigma_{0a}\theta_a)\psi\end{gather*}
where $\theta_a$ are transformation parameters and $\Sigma_{0a}$ are
matrices  (\ref{n2}).

In particular, the  spinors $\psi^\varepsilon_\nu$ defined by
relation (\ref{6}) are transformed as:
\begin{gather}\la{infi}\psi^\varepsilon_\nu\to
\left(1-\frac{\varepsilon
p_a\theta_a}{2p}\right)\psi^\varepsilon_\nu+\frac{\ri
p_0S_{ab}p_b\theta_a}{p}
\left(\frac{p_0}p\psi^\varepsilon_\nu+\left(\frac{p_0}p+
\nu\right)(\delta_{\varepsilon\nu}\psi^\alpha_\alpha-\psi^\nu_\varepsilon)\right)
.\end{gather}

 Starting with (\ref{n1}) it is possible to find also finite
Lorentz transformations. To do it it is sufficient to solve the Lie
equations for transformations generated by these infinitesimal
operators. It is sufficient to restrict ourselves to the particular
case $\theta_1=\theta_2=0, \ \theta_3=\theta$ then the generic
Lorenz boost can be obtained by a rotation transformation. In this
particular case  the Lorentz boost  generated by infinitesimal
operator (\ref{n1}) has the following form:
\begin{gather}\la{Lie4}\begin{split}&p_1'=p_1,\quad p_2'=p_2,\\&p_3'=p_3\cosh\theta-p_0\sinh\theta,\\&
p_0'=p_0\cosh\theta-p_0\sinh\theta,\end{split}\\
\la{Lie41}\begin{split}&\psi'^\varepsilon_\nu
=\left(F(\theta)(A_++\ri
BS_{3a}p_a)+\sinh\frac\theta2\frac\nu{p}(\ri A_-S_{3a}p_a-\tilde
p^2B)\right)\psi^\varepsilon_\nu\\&+\left(F(\theta)(A_-+\ri
BS_{3a}p_a)+\sinh\frac\theta2\frac\nu{p}(\ri A_+S_{3a}p_a-\tilde
p^2B)\right)\gamma_5
(\delta_{\varepsilon\nu}\psi^\alpha_\alpha-\psi^\nu_\varepsilon)\end{split}
\end{gather}
where
\begin{gather*}F(\theta)=\cosh\frac\theta2-
\sinh\frac\theta2\frac{\varepsilon p_3}p,\quad A_\pm=\frac{\tilde
p^2+p_3p_3'\pm pp'}{2pp'}, \quad B=\frac{p_3'-p_3}{2pp'},\\\tilde
p^2=p_1^2+p_2^2, \quad p=\sqrt{p_1^2+p_2^2+p_3^2},\quad
p'=\sqrt{p'^2_1+p'^2_2+p'^2_3}.\end{gather*}

We will prove rather complicated formulae (\ref{Lie41}) and
(\ref{infi}) in Appendix.

\section{One more  non-standard Dirac equation}

Let us return to equations  (\ref{4}). We will consider  their
solutions as functions of four independent variables $p_0, p_1, p_2$
and $p_3$ which are equal in rights.  Then we postulate invariance
of (\ref{4}) with respect to the following discrete transformations
 \begin{gather*}
\Psi(p_0,{\bf p})\to P\Psi(p_0,{\bf p})=\gamma_0\Psi(p_0,-{\bf
p}),\\  \Psi(p_0,{\bf p})\to  T\Psi(p_0,{\bf
p})=\gamma_1\gamma_3\Psi^*(-p_0,{\bf p}),\\\Psi(p_0,{\bf p})\to C
\Psi(p_0,{\bf p})=\gamma_2\Psi^*(p_0,{\bf p}).
\end{gather*}

By definition, $P$ commutes with $\gamma^\mu p_\mu$ while $C$ and
$T$ anticommute with this term. Thus, in order to equation (\ref{4})
be invariant with respect to these transformations, it is necessary
to ask for the following conditions for  $I$:
\[PI=IP,\quad CI=-IC,\quad TI=-IT.\]
In addition,  to guarantee  correct dispersion relations (\ref{7}),
(pseudo)involution $I$ should satisfy one of the following relation:
\begin{gather}\la{23}\gamma^\mu p_\mu I=I\gamma^\mu p_\mu, \quad I^2=1\end{gather}
or, alternatively,
\begin{gather}\la{24}\gamma^\mu p_\mu I=-I\gamma^\mu p_\mu, \quad I^2=-1.\end{gather}

These conditions together with the requirement of Lorenz invariance
leave the only possibility for $I$, i.e., $I=i\gamma_5PT$. In this
case equation (\ref{4}) takes the following form:
\begin{gather}\la{n11}\left(\gamma^\mu p_\mu+im\gamma_5PT\right)\Psi=0\end{gather}
where we change $\psi\to\Psi$ to discriminate solutions of (\ref{n11})
from wave functions discussed in the previous sections.

Operator $i\gamma_5PT$ commutes with $\gamma^\mu p_\mu$ and is an
involution, i.e., $(i\gamma_5PT)^2=1$. Thus acting on equation
(\ref{n11}) from the left by $\left(\gamma^\mu
p_\mu-im\gamma_5PT\right)$ we immediately find that equation
(\ref{n11}) generates condition (\ref{7}). Moreover, in contrast with
(\ref{10}), equation (\ref{n11}) is transparently relativistic
invariant.

Let show that there exist some intriguing  similarities between
solutions of equation (\ref{n11}) and ELKO. Indeed, acting to this
equation from the left by the projector
\[\hat P^\varepsilon_\lambda=\frac14(1+\varepsilon C)(1+\lambda \gamma_5PCT)\]
 and using the identities
\[\begin{split}& \hat P^\varepsilon_\lambda\gamma^\mu p_\mu=\gamma^\mu p_\mu\hat P^{-\varepsilon}_{-\lambda},\quad \hat P^\varepsilon_\lambda i\gamma_5PT=i\varepsilon \lambda\hat P^{-\varepsilon}_\lambda,\end{split}\]
one can make sure  that the linearly independent functions
\begin{gather}\la{18}\Psi^+_+=\hat P^+_+\Psi,\ \Psi^+_-=\hat P^+_-\Psi,\ \Psi^-_-=\hat P^-_-\Psi,\ \Psi^-_+=\hat P^-_+\Psi\end{gather}
satisfy the fundamental equations of ELKO theory, given by formulae
(\ref{1}).

In accordance with (\ref{18}), spinors $\Psi^\varepsilon_\lambda$
satisfy the following conditions:
\begin{gather}\la{19}C\Psi^\varepsilon_\lambda=\varepsilon
\Psi^\varepsilon_\lambda,\quad
\gamma_5PCT\Psi^\varepsilon_\lambda=\lambda\Psi^\varepsilon_\lambda.\end{gather}
Since   operator $\gamma_5PCT$ is nothing but a total reflection of
all independent variables, the latter equation can be rewritten in
the following form:
\begin{gather}\la{20}\Psi^\varepsilon_\lambda(-p_0,-{\bf p})=\lambda \Psi^\varepsilon_\lambda(p_0,{\bf p}).\end{gather}

Thus, like  ELKO, functions  (\ref{18}) are eigenvectors of the
charge conjugation operator, satisfying equations (\ref{1}) and
(\ref{2}). However, in contrast with (\ref{2a}), they are not
eigenvectors  of the chirality operator, but are eigenvectors of
$\gamma_5PCT$ instead.

The fundamental distinction of the introduced spinors $\Psi$ from
 ELKO is that, in contrast with (\ref{2a}), both conditions
(\ref{19}) are transparently relativistic invariant.

Finally, let us represent a non-standard Dirac equation in
configuration space:
\begin{gather}\la{22}(i\gamma^\mu\p_\mu-imR)\Psi(x)=0\end{gather}
where $x=(x_0, x_1, x_2, x_3)$ and $R$ is the total reflection
operator whose action on $\Psi(x)$ is defined as
$R\Psi(x)=\Psi(-x)$.

Acting on (\ref{22}) from the left by projectors $P_+=\frac12(1+R)$
and  $P_-=\frac12(1-R)$, we obtain the following system
\begin{gather}\la{23a}\begin{split}&i\gamma^\mu\p_\mu\Psi_+(x)=im\Psi_-(x),\\
&i\gamma^\mu\p_\mu\Psi_-(x)=-im\Psi_+(x)\end{split}
\end{gather}
where $\Psi_\pm=P_\pm\Psi$ are eigenvectors of the total reflection
operator.

Up to the meaning of vectors $\Psi_\pm$ the system (\ref{23a})
coincides with equations for ELKO in configuration space, presented,
e.g., in \cite{Ahlu1}.
\section{Discussion}

In this paper  a new look on the kinematical grounds of the ELKO
theories is presented. Namely, we give  a compact four component formulation (\ref{5})
of kinematic equation for these spinors. Then, a simple and
straightforward connection between Dirac spinors and ELKO is
presented. Finally, the transformation properties of ELKO w.r.t.
Lorentz boost are discussed. Since ELKO also satisfy equations
(\ref{1}), (\ref{2}) and (\ref{2a}) by construction, the results of
the present paper could be interesting for experts in ELKO approach.

The transformations connecting the Dirac spinors and ELKO were
studied in \cite{RohRoh}. However, these transformations where
made under the supposition that the left handed components of the
Dirac and ELKO coincide. In order  this supposition to be correct,
the Dirac spinors should satisfy one of the additional constraints
discussed in \cite{RohRoh}. The transformation for ELKO
generated by operator (\ref{11}) is valid without additional
constraints.

We show that in spite of that the eigenvectors of dual helicity
operator are not  covariant subjects, ELKO  form a carrier space
of the representation of Poincar\'e  group, whose generators are
given by equations (\ref{14}).
 The corresponding boost
generators  do not have a covariant form. Nevertheless, they
generate covariant transformations for the case when the new frame
of reference moves parallel to particle momentum. These facts can be
used for justification of ELKO approach which is appears to be
non-covariant in the standard meaning \cite{crash}.

Equations, presented in Section 6 are just toy models which are
seemed to be rather peculiar. In particular, the equality in rights
of all variables in equation (\ref{n11}) is a natural but
non-standard proposition. Usually $p_0$ is considered as a
distinguished variable which is not affected by the time reflection.

 The formal analogy of these equations with kinematic equations for
ELKO is rather curious. And this analogy generates  a challenge to
search  for possible applications of the corresponding fields in
non-standard physical theories.

   A specific feature of equations (\ref{5}), (\ref{n11}) and (\ref{22}) is
   that they include  involutions $C\sigma_p$, $\gamma_5PCT$ or $R$ as
   essential constructive elements. Such (and other) involutions  present
   additional tools for creating   alternatives to Dirac's factorization of
   the Klein-Gordon equation. Apparently the first  example of such
   non-standard factorization was proposed long time ago in paper
   \cite{bied} where a two component version of first order equations for
   a massive spinor field was discussed.

   It is interesting to note that equation (\ref{10}) for ELKO can be
   decoupled to two subsystems each of which, like
   equation proposed in \cite{bied}, is two-component like
   equation proposed in \cite{bied}. Indeed, hamiltonian $H$
   commutes with diagonal matrix $\gamma_5$, and so
   \begin{gather*}H=\begin{pmatrix}H_+&0\\0&H_-\end{pmatrix}\end{gather*}
   where
   \begin{gather*}H_\pm=\sigma_ap_a\left(\pm1-
   \frac{im}p\sigma_2\kappa\right)\end{gather*}
   and $\sigma_a$ are Pauli matrices. Let us note that boost
   transformations mix eigenvectors of $H_+$ and $H_-$.

   Involutive discrete symmetries  have useful applications in
   construction of exact Foldy-Wouthuysen transformations \cite{N3} and
   generating of non-standard realizations of symmetry algebras and
   superalgebras \cite{N4}, \cite{N5}, \cite{N6},
   \cite{N2}, \cite{zhe}. We see  that  such involutions
   can also be effectively used  to formulate a compact equation for ELKO.

   \renewcommand{\theequation}{A\arabic{equation}} %
\setcounter{equation}{0}
\section{Appendix. Integration  of Lie equations}
The boost generator (\ref{n1}) includes involutions $C$ and
$\sigma_p$, and the corresponding Lie equations are rather
complicated. We will find the action of this generator on
eigenvectors $\psi^\varepsilon_\nu$ of these involutions whose
formal definition is given by equation (\ref{6}). Let us define the
corresponding matrix entries of boost operator (\ref{n1}):
\begin{gather}\la{J0a}(J_{0a}')^{\varepsilon\varepsilon'}_{\nu\nu'}=
(K_a)^{\varepsilon\varepsilon'}_{\nu\nu'}+(\hat
S_{0a})^{\varepsilon\varepsilon'}_{\nu\nu'}\end{gather} where
\begin{gather}\la{DW}(K_a)^{\varepsilon\varepsilon'}_{\nu\nu'}=
P^\varepsilon_\nu K_a P^{\varepsilon'}_{\nu'},\qquad (\hat
S_{0a})^{\varepsilon\varepsilon'}_{\nu\nu'}=P^\varepsilon_\nu \hat
S_a P^{\varepsilon'}_{\nu'}.
\end{gather}
Then action of the boost generator on $\psi^\varepsilon_\nu$ can be
represented as:
\begin{gather}\la{Act}J_{0a}'\psi^\varepsilon_\nu=
(K_a)^{\varepsilon\varepsilon'}_{\nu\nu'}\psi^{\varepsilon'}_{\nu'}+(\hat
S_{0a})^{\varepsilon\varepsilon'}_{\nu\nu'}\psi^{\varepsilon'}_{\nu'}.\end{gather}
where
\begin{gather}\la{entri}\begin{split}&
(K_a)^{\varepsilon\varepsilon'}_{\nu\nu'}=-p_0\frac{\p}{\p
p_a}\delta_{\varepsilon\varepsilon'} \delta_{\nu\nu'}+\frac{2\ri
p_0}{p^2}S_{ab}p_b M^{\varepsilon\varepsilon'}_{\nu\nu'},\\&
(S_{0a})^{\varepsilon\varepsilon'}_{\nu\nu'}= -\frac{\varepsilon
p_a}{2p} \delta_{\varepsilon\varepsilon'}
\delta_{\nu\nu'}+\frac{\ri\nu}p\gamma_5S_{ab}p_b(2
M^{\varepsilon\varepsilon'}_{\nu\nu'}-\delta_{\varepsilon\varepsilon'}
\delta_{\nu\nu'})
\end{split}\end{gather} and
\begin{gather}\la{MMM}M^{\varepsilon\varepsilon'}_{\nu\nu'}=\frac12(
\delta_{\varepsilon\varepsilon'} \delta_{\nu\nu'}
+(\delta_{\varepsilon\nu}
\delta_{\varepsilon'\nu'}-\delta_{\varepsilon\nu'}
\delta_{\varepsilon'\nu})\gamma_5).\end{gather} The expressions
(\ref{entri}) for the entries of the boost generator can be
calculated directly using definitions (\ref{P}), (\ref{Ka}) and
(\ref{DW}). In particular case $a=3$ these expressions  are reduced
to the following form:
\begin{gather}\la{entri2}
(K_3)^{\varepsilon\varepsilon'}_{\nu\nu'}=-p_0\frac{\p}{\p
p_3}\delta_{\varepsilon\varepsilon'} \delta_{\nu\nu'}+\frac{2\ri
p_0}{p^2}(S_{31}p_1+S_{32}p_2)M^{\varepsilon\varepsilon'}_{\nu\nu'}
,\\\la{entri3}(S_{03})^{\varepsilon\varepsilon'}_{\nu\nu'}=
-\frac{\varepsilon p_3}{2p} \delta_{\varepsilon\varepsilon'}
\delta_{\nu\nu'}+\frac{\ri\nu}p\gamma_5(S_{31}p_1+S_{32}p_2) (2
M^{\varepsilon\varepsilon'}_{\nu\nu'}-\delta_{\varepsilon\varepsilon'}
\delta_{\nu\nu'}) .\end{gather}

Since operators $ (K_a)^{\varepsilon\varepsilon'}_{\nu\nu'}$ and
$(\hat S_{0a})^{\varepsilon\varepsilon'}_{\nu\nu'}$ commute each
other, the finite boost transformations produced by generator
(\ref{J0a}) can be represented as a product of two transformations
generated by these commuting parts of $
(J_a)^{\varepsilon\varepsilon'}_{\nu\nu'}$:
\begin{gather}\la{tran}p_\mu\to p_\mu'=\Lambda_{\mu\nu}p^\nu,\quad
\psi^\varepsilon_\nu\to\psi'^\varepsilon_\nu\end{gather} and
\begin{gather}\la{trans}p_\mu'\to p_\mu''=p_\mu',\quad
{\psi'}^\varepsilon_\nu\to{\psi''}^\varepsilon_\nu\end{gather} where
(\ref{tran}) and (\ref{trans}) are the generic forms of
transformation generated by $
(K_a)^{\varepsilon\varepsilon'}_{\nu\nu'}$ and $(\hat
S_{0a})^{\varepsilon\varepsilon'}_{\nu\nu'}$ correspondingly,
$\Lambda^\mu_\nu$ is a Lorentz transformation matrix.

To find the finite transformations of $\tilde\psi^\varepsilon_\nu$
generated by the infinitesimal operator (\ref{entri2}) it is
sufficient to solve the following Lie equations:

\begin{gather}\la{Lie1}\begin{split}&\frac{\p p'_3}{\p
\theta}=-E',\\
&p'_3|_{\theta=0}=p_3;\end{split}\end{gather}
\begin{gather}\la{Lie2}\begin{split}&\frac{\p {\psi'^\lambda_\mu}}
{\p \theta}=\frac{\ri \tilde p
p'_0}{p'^2}\Lambda^{\lambda\varepsilon'}_{\mu\nu'}
{\psi'^{\varepsilon'}_{\nu'}},\\&
{\psi'^\lambda_\mu}|_{\theta=0}=\psi^\lambda_\mu\end{split}\end{gather}
where
\begin{gather}\la{La}\Lambda^{\lambda\varepsilon'}_{\mu\nu'}=\frac2{\tilde p}
(S_{31}p_1+S_{32}p_2)M^{\lambda\varepsilon'}_{\mu\nu'}\end{gather}
and $\tilde p=\sqrt{p_1^2+p_2^2}$.

 Since operator (\ref{entri2}) does
not include derivations w.r.t. $p_1$ and $p_2$, these variables are
kept unchanged, i.e., $p'_1=p_1$ and $p'_2=p_2$. Thus
${p'}^2={p'}^2_1+{p'}^2_2+{p'}^2_3=p^2_1+p^2_2+{p'}^2_3.$

Equations (\ref{Lie1}) and (\ref{Lie2}) can be integrated in closed
form and give the following transformed functions:
\begin{gather}\psi'^\lambda_\mu =\psi^{\lambda}_\mu+\left(\frac{
p^2+\left(2\ri(S_{31}p_1+S_{32}p_2)+p_3\right)(p_3'-p_3)}{pp'}-1\right)
M^{\lambda\varepsilon'}_{\mu\nu'}\psi^{\varepsilon'}_{\nu'}
\la{Lie5}.\end{gather}

To integrate (\ref{Lie2}) we use the following observations:
\begin{itemize}
\item The $\theta$-dependent multiplier in the r.h.s. of the first
equation (\ref{Lie2}) can be represented as:
\begin{gather*}\frac{p_0'}{p'^2}=\frac{\frac{\partial
p_3'}{\partial \theta}}{\tilde p^2+p_3'^2};\end{gather*}
\item Matrix $\Lambda$ whose entries are given by equation
(\ref{La}) does not depend on $\theta$ and satisfies the following
conditions:
\begin{gather*} \Lambda^2=M,\quad \Lambda^3=\Lambda\end{gather*}
where $M$ is the matrix with entries (\ref{MMM}).
\end{itemize}

Thus we find explicitly the finite transformation (\ref{tran})
generated by $(K_3)^{\lambda\varepsilon'}_{\mu\nu'}$. The next step
is to find transformation (\ref{trans}) generated by
$(S_{03})^{\varepsilon\lambda}_{\nu\mu}$. The corresponding Lie
equations reads:
\begin{gather}\la{Lie3}\begin{split}&\frac{\p
{\psi''^\varepsilon_\nu}} {\p
\theta}=(S_{03})^{\varepsilon\lambda}_{\nu\mu}\psi'^\lambda_\mu,\\&
{\psi''^\lambda_\mu}|_{\theta=0}=\psi'^\lambda_\mu\end{split}\end{gather}
and have the following solution:
\begin{gather}\la{Lie7}\psi''^\varepsilon_\nu=\left(\cosh\frac\theta2-\sinh\frac\theta2\frac{\varepsilon
p_3}{p}\right)\delta_{\varepsilon\lambda}\delta_{\nu\mu}+\ri\sinh\frac\theta2\nu
S_{3a}p_a\gamma_5(\delta_{\varepsilon\nu}
\delta_{\lambda\mu}-\delta_{\varepsilon\mu}
\delta_{\lambda\nu}).\end{gather} The completed boost transformation
is a product of transformations (\ref{Lie5}) and (\ref{Lie7}).
Substituting the expression (\ref{Lie5}) for $\psi'^\lambda_\mu$
into (\ref{Lie7}) we obtain equation (\ref{Lie41}).

\end{document}